\documentclass
[prl,10pt,letterpaper,twocolumn,showpacs,bibnotes,notitlepage,final,balancelastpage,groupedaddress]{revtex4}%
\usepackage{amssymb}
\usepackage{amsmath}
\usepackage{amsfonts}
\usepackage{graphicx}%
\begin{document}
\title{Violation of Locality Without Inequalities for Multiparticle Perfect Correlations}
\author{Zeng-Bing Chen}
\email{zbchen@ustc.edu.cn}
\affiliation{Department of Modern Physics, University of Science and Technology of China,
Hefei, Anhui 230027, China}
\author{Nai-Le Liu}
\affiliation{Department of Modern Physics, University of Science and Technology of China,
Hefei, Anhui 230027, China}
\author{Sixia Yu}
\affiliation{Department of Modern Physics, University of Science and Technology of China,
Hefei, Anhui 230027, China}
\author{Yong-De Zhang}
\affiliation{Department of Modern Physics, University of Science and Technology of China,
Hefei, Anhui 230027, China}
\date{\today }

\begin{abstract}
We prove that for a three-qubit system in the Greenberger-Horne-Zeilinger
(GHZ) state, locality \textit{per se} is in conflict with the perfect
GHZ\ correlations. The proof does not in any way use the realism assumption
and can lead to a refutation of locality. We also provide inequalities that
are imposed by locality. The experimental confirmation of the present
reasoning may imply a genuine quantum nonlocality and will deepen our
understanding of nonlocality of nature.

\end{abstract}
\maketitle

Bell's discovery of his celebrated inequalities (or more generally, Bell's
theorem) \cite{Bell,CHSH,Bell-book}, which are derived from Einstein,
Podolsky, and Rosen's notion of local realism \cite{EPR}, is recognized as
\textquotedblleft one of the profound scientific discoveries of the
century\textquotedblright\ \cite{Aspect}. Bell's inequalities have already
been generalized to the multiparticle cases \cite{Mermin65,Bell-n}. Bell's
theorem with inequalities demonstrates the conflict between local realism and
quantum mechanics (QM) for certain \textit{statistical predictions} of quantum
theory.\ Bell's theorem without inequalities has also been demonstrated for
multiparticle Greenberger-Horne-Zeilinger (GHZ) states
\cite{GHZ-89,GHZ-90,Mermin} and for nonmaximally entangled biparticle states
\cite{Hardy}. Recently, a GHZ-type theorem was proved for the case of two
pairs of maximally entangled qubits \cite{Cabello} and for the case of two
doubly-entangled photons \cite{Chen-GHZ}. Strikingly, for Bell's theorem
without inequalities the contradiction between QM and local realism is of a
non-statistical nature. This is known as the \textquotedblleft all versus
nothing\textquotedblright\ proof of Bell's theorem.

Locality and realism are two indispensable assumptions
\cite{Werner-rev,assume} that are used in Bell's theorem demonstrating the
contradictions between QM and local realism. So far, many experiments
\cite{Aspect,loophole} completely confirmed the statistical predictions of QM
against local realistic theories though some loopholes still remain to be
closed \cite{loophole}. Experimental tests of the GHZ theorem \cite{Pan-GHZ}
and Hardy's theorem \cite{Hardy-exp} have also been performed, confirming
again the correctness of QM. These experiments to test Bell's theorem rule out
local realism and necessarily imply that at least one of the two assumptions
(realism or locality) should be incompatible with QM \cite{Werner-rev,assume}.

However, since both realism \textit{and} locality are necessary ingredients in
Bell's theorem, all experiments designed to test Bell's theorem (with and
without inequalities) rule out neither locality nor realism separately.
Logically, a new falsifiable formulation beyond Bell's theorem is thus
desirable to test QM versus realism and versus locality separately. Very
recently, this kind of formulation has been suggested such that the separate
role of the locality or realism assumption can be tested for statistical
predictions of QM \cite{Chen}. In particular, we derived for the Bell
experiments two testable inequalities by considering realism or locality
separately. This result can lead to experimental tests of locality of nature
beyond Bell's theorem. In Ref. \cite{Chen}, however, only statistical
predictions of QM were considered. In QM, there are also non-statistical
(i.e., definite) predictions, or perfect correlations, such as those
considered in Bell's theorem without inequalities
\cite{GHZ-89,GHZ-90,Mermin,Hardy,Cabello,Chen-GHZ}. It is thus important to
see if one can test locality of nature for these definite predictions of QM.

In this Letter we prove that for a three-qubit system in the GHZ state
locality is indeed in conflict with the perfect correlations predicted by QM.
To this end, one needs to adopt a locality condition without resorting to any
specific theory (quantum or realistic). Then it is shown that the definite
predictions as used in the usual GHZ reasoning \cite{GHZ-89,GHZ-90,Mermin} are
incompatible with the locality condition. The proof thus does not in any way
use the realism assumption or other counterfactual reasonings \cite{Stapp-GHZ}.

To warm up and for comparison, let us first recall the standard reasoning of
the GHZ theorem \cite{GHZ-89,GHZ-90,Mermin,Pan-GHZ}. Consider the three
qubits, each of which is spacelike separated from the remaining qubits, in the
GHZ state
\begin{equation}
\left\vert \mathrm{\Delta}\right\rangle =\frac{1}{\sqrt{2}}\left(  \left\vert
\uparrow\right\rangle _{1}\left\vert \uparrow\right\rangle _{2}\left\vert
\uparrow\right\rangle _{3}+\left\vert \downarrow\right\rangle _{1}\left\vert
\downarrow\right\rangle _{2}\left\vert \downarrow\right\rangle _{3}\right)  ,
\label{ghz}%
\end{equation}
where $\left\vert \uparrow\right\rangle \equiv\left\vert +1\right\rangle
$\ and $\left\vert \downarrow\right\rangle \equiv\left\vert -1\right\rangle
$\ are two orthonormal states (e.g., the spin-up and spin-down states) of a
qubit. Following the GHZ argument \cite{GHZ-89,GHZ-90,Mermin,Pan-GHZ}, it is
directly verifiable that the GHZ state (\ref{ghz}) satisfies the eigenvalue
equations as
\begin{align}
\sigma_{1x}\sigma_{2x}\sigma_{3x}\left\vert \mathrm{\Delta}\right\rangle  &
=\left\vert \mathrm{\Delta}\right\rangle ,\ \ \sigma_{1x}\sigma_{2y}%
\sigma_{3y}\left\vert \mathrm{\Delta}\right\rangle =-\left\vert \mathrm{\Delta
}\right\rangle ,\nonumber\\
\sigma_{1y}\sigma_{2x}\sigma_{3y}\left\vert \mathrm{\Delta}\right\rangle  &
=-\left\vert \mathrm{\Delta}\right\rangle ,\ \ \sigma_{1y}\sigma_{2y}%
\sigma_{3x}\left\vert \mathrm{\Delta}\right\rangle =-\left\vert \mathrm{\Delta
}\right\rangle . \label{ghz4}%
\end{align}
Due to the perfect GHZ correlations, one can determine the value of any
observable ($\sigma_{1x}$, $\sigma_{1y}$, $\sigma_{2x}$,$\ldots$) for a qubit
by performing appropriate measurements on the other two qubits. This allows
one to establish a local realistic interpretation of the quantum-mechanical
results (\ref{ghz4}), i.e., one assumes that the \textit{individual} value of
any operator ($\sigma_{1x}$, $\sigma_{1y}$, $\sigma_{2x}$,$\ldots$) is
predetermined, e.g., by local hidden variables, regardless of any set of
measurements on the three qubits. However, the GHZ theorem states that such a
local realistic interpretation is always incompatible with the non-statistical
quantum predictions (\ref{ghz4}).

In the GHZ reasoning, the realism and locality assumptions are explicitly used
to establish the confliction with QM. By realism one means that for any
physical quantity, there is a definite value predetermined possibly by certain
classical hidden variables. Meanwhile, locality requires that for any physical
theory (quantum or classical), the experimental results obtained from a
physical system at one location should be independent of any observations or
actions made at any other spacelike separated locations. Local realism in the
context of the GHZ experiment has been ruled out by recent experiments
\cite{Pan-GHZ}. We notice that Stapp \cite{Stapp-GHZ} made an alternative
argument of the GHZ experiment. However, though the argument does not use
realism explicitly, it resorts to some other counterfactual reasonings, which
remain controversial \cite{Stapp-comment}.

The Pauli operators $\sigma_{1x}$ and $\sigma_{1y}$ for qubit-$1$ can be
represented by $\sigma_{1x}=\sum_{i_{x}=\pm1}i_{x}\left\vert i_{x}%
\right\rangle \left\langle i_{x}\right\vert $ and\ $\sigma_{1y}=\sum
_{i_{y}=\pm1}i_{y}\left\vert i_{y}\right\rangle \left\langle i_{y}\right\vert
$, where $\left\vert i_{x}\right\rangle =\frac{1}{\sqrt{2}}\left(  \left\vert
\uparrow\right\rangle _{1}+i_{x}\left\vert \downarrow\right\rangle
_{1}\right)  $\ [$\left\vert i_{y}\right\rangle =\frac{1}{\sqrt{2}}\left(
\left\vert \uparrow\right\rangle _{1}+ii_{y}\left\vert \downarrow\right\rangle
_{1}\right)  $] are the eigenstates of $\sigma_{1x}$ ($\sigma_{1y}$) with
eigenvalues $i_{x}=\pm1$ ($i_{y}=\pm1$). The corresponding Pauli operators for
the remaining qubits can be similarly represented, but with $i_{x,y}$ being
replaced by $j_{x,y}=\pm1\ $for qubit-$2$ and $k_{x,y}=\pm1\ $for qubit-$3$.
In terms of these eigenstates, the GHZ state (\ref{ghz}) also reads$\ $%
\begin{equation}
\left\vert \mathrm{\Delta}\right\rangle =\sum_{i_{x},j_{x},k_{x}=\pm1}%
\Psi_{i_{x},j_{x},k_{x}}\left\vert i_{x},j_{x},k_{x}\right\rangle ,
\label{ghz-exp}%
\end{equation}
where $\Psi_{i_{x},j_{x},k_{x}}$ are the probability amplitudes in the
eigenstate base. The first eigen-equation in (\ref{ghz4}) can then be
rewritten as
\begin{equation}
\sum_{i_{x},j_{x},k_{x}=\pm1}i_{x}j_{x}k_{x}\Psi_{i_{x},j_{x},k_{x}}\left\vert
i_{x},j_{x},k_{x}\right\rangle =\left\vert \mathrm{\Delta}\right\rangle ,
\label{ghzi}%
\end{equation}
which, together with Eq. (\ref{ghz-exp}), gives immediately
\begin{equation}
\sum_{i_{x},j_{x},k_{x}=\pm1}i_{x}j_{x}k_{x}\left\vert \Psi_{i_{x},j_{x}%
,k_{x}}\right\vert ^{2}=1. \label{ijk1}%
\end{equation}
Standard QM states that $\left\vert \Psi_{i_{x},j_{x},k_{x}}\right\vert ^{2} $
are the probabilities of measuring $\sigma_{1x}$\ with the outcome $i_{x}$,
$\sigma_{2x}$\ with the outcome $j_{x}$, and $\sigma_{3x}$\ with the outcome
$k_{x}$. Similarly, one obtains
\begin{align}
\sum_{i_{x},j_{y},k_{y}=\pm1}i_{x}j_{y}k_{y}\left\vert \Psi_{i_{x},j_{y}%
,k_{y}}\right\vert ^{2}  &  =-1,\label{ijk2}\\
\sum_{i_{y},j_{x},k_{y}=\pm1}i_{y}j_{x}k_{y}\left\vert \Psi_{i_{y},j_{x}%
,k_{y}}\right\vert ^{2}  &  =-1,\label{ijk3}\\
\sum_{i_{y},j_{y},k_{x}=\pm1}i_{y}j_{y}k_{x}\left\vert \Psi_{i_{y},j_{y}%
,k_{x}}\right\vert ^{2}  &  =-1, \label{ijk4}%
\end{align}
where the probabilities $\left\vert \Psi_{i_{x},j_{y},k_{y}}\right\vert ^{2}
$, $\left\vert \Psi_{i_{y},j_{x},k_{y}}\right\vert ^{2}$, and $\left\vert
\Psi_{i_{y},j_{y},k_{x}}\right\vert ^{2}$\ are to be understood similarly to
$\left\vert \Psi_{i_{x},j_{x},k_{x}}\right\vert ^{2}$. Note that
(\ref{ijk1}-\ref{ijk4}) are derived using only standard QM.

Since $\left\vert i\right\vert =\left\vert j\right\vert =\left\vert
k\right\vert =1$\ (Note that $\sigma_{x}^{2}=\sigma_{y}^{2}=I$), for nonzero
$\left\vert \Psi_{i,j,k}\right\vert ^{2}$ ($\leq1$) Eqs. (\ref{ijk1}%
-\ref{ijk4}) give immediately
\begin{align}
i_{x}j_{x}k_{x}  &  =1,\ \ \ i_{x}^{\prime}j_{y}k_{y}=-1,\nonumber\\
i_{y}j_{x}^{\prime}k_{y}^{\prime}  &  =-1,\ \ \ i_{y}^{\prime}j_{y}^{\prime
}k_{x}^{\prime}=-1, \label{allno}%
\end{align}
which are actually the measured results of a single run of the GHZ experiment.
However, the relations in (\ref{allno}) cannot lead to the usual GHZ conflict
except that there is a logic link ensuring $\gamma_{x}=\gamma_{x}^{\prime}$
and $\gamma_{y}=\gamma_{y}^{\prime}$ (with $\gamma=i,j,k$). The realism
assumption underlying the GHZ reasoning \cite{GHZ-89,GHZ-90,Mermin,Pan-GHZ} or
Stapp's counterfactual reasoning \cite{Stapp-GHZ} can provide such a logic
link that gives an all-versus-nothing conflict for only a single run of the
GHZ experiment.

By contrast to the GHZ/Stapp reasoning, here we \textit{assume only locality}.
Then how the above three-qubit GHZ correlations can be interpreted by a
localist? According to local causality, events occurring in the backward light
core of a qubit may affect the events occurring on the qubit. Thus, only
events in the overlap of the backward light cores of the three qubits may be
\textquotedblleft common causes\textquotedblright\ \cite{Bell-book} of the
events occurring on the three qubits. The localist may then interpret the GHZ
correlations being solely coming from the common causes that are certain
physical events. Similarly to Ref. \cite{Chen}, the locality condition in the
present three-qubit case reads
\begin{equation}
\left\vert \Psi_{i,j,k}\right\vert ^{2}=\sum_{\mu}p_{\mu}p_{i}^{\mu}p_{j}%
^{\mu}p_{k}^{\mu}, \label{local4}%
\end{equation}
where $p_{i,j,k}^{\mu}$\ are local probabilities (e.g., $p_{i_{x}}^{\mu}$
represents the probability of measuring $\sigma_{1x}$\ with the outcome
$i_{x}$) conditioned on a given common cause\ $\mu$; $p_{\mu}$ ($\geq0$) are
the probabilities for the given cause $\mu$ to occur, and $\sum_{\mu}p_{\mu
}=1$. The locality condition (\ref{local4}) for the GHZ state (\ref{ghz}) has
been imposed in such a way that the given common cause can affect all local
probabilities; conditioned on the same common cause, observable probabilities
for the qubits must be mutually independent.

Using the locality condition (\ref{local4}) and defining
\begin{align}
\overline{i}_{x}^{\mu}  &  \equiv\sum_{i_{x}=\pm1}i_{x}p_{i_{x}}^{\mu
},\ \ \overline{i}_{y}^{\mu}\equiv\sum_{i_{y}=\pm1}i_{y}p_{i_{y}}^{\mu
},\nonumber\\
\bar{j}_{x}^{\mu}  &  \equiv\sum_{j_{x}=\pm1}j_{x}p_{j_{x}}^{\mu},\ \ \bar
{j}_{y}^{\mu}\equiv\sum_{j_{y}=\pm1}j_{y}p_{j_{y}}^{\mu},\nonumber\\
\bar{k}_{x}^{\mu}  &  \equiv\sum_{k_{x}=\pm1}k_{x}p_{k_{x}}^{\mu},\ \ \bar
{k}_{y}^{\mu}\equiv\sum_{k_{y}=\pm1}k_{x}p_{k_{y}}^{\mu}, \label{def-ijk}%
\end{align}
Eqs. (\ref{ijk1}-\ref{ijk4}) then become
\begin{align}
\sum_{\mu}p_{\mu}\overline{i}_{x}^{\mu}\cdot\bar{j}_{x}^{\mu}\cdot\bar{k}%
_{x}^{\mu}  &  =+1,\ \ \sum_{\mu}p_{\mu}\overline{i}_{x}^{\mu}\cdot\bar{j}%
_{y}^{\mu}\cdot\bar{k}_{y}^{\mu}=-1,\nonumber\\
\sum_{\mu}p_{\mu}\overline{i}_{y}^{\mu}\cdot\bar{j}_{x}^{\mu}\cdot\bar{k}%
_{y}^{\mu}  &  =-1,\ \ \sum_{\mu}p_{\mu}\overline{i}_{y}^{\mu}\cdot\bar{j}%
_{y}^{\mu}\cdot\bar{k}_{x}^{\mu}=-1, \label{confl}%
\end{align}
which, for nonzero $p_{\mu}$ ($\leq1$) and any given $\mu$, lead to
\begin{align}
\overline{i}_{x}^{\mu}\bar{j}_{x}^{\mu}\bar{k}_{x}^{\mu}  &
=1,\ \ \ \overline{i}_{x}^{\mu}\bar{j}_{y}^{\mu}\bar{k}_{y}^{\mu
}=-1,\nonumber\\
\overline{i}_{y}^{\mu}\bar{j}_{x}^{\mu}\bar{k}_{y}^{\mu}  &
=-1,\ \ \ \overline{i}_{y}^{\mu}\bar{j}_{y}^{\mu}\bar{k}_{x}^{\mu}=-1.
\label{confl2}%
\end{align}
Obviously, $\left\vert \overline{i}_{x,y}^{\mu}\right\vert =1$, $\left\vert
j_{x,y}^{\mu}\right\vert =1$, and $\left\vert k_{x,y}^{\mu}\right\vert =1$.
Similarly to the GHZ argument, these relations are also not mutually
consistent: The latter three relations in Eq. (\ref{confl2}) give
$\overline{i}_{x}^{\mu}\bar{j}_{x}^{\mu}\bar{k}_{x}^{\mu}\left(  \overline
{i}_{y}^{\mu}\right)  ^{2}\left(  \bar{j}_{y}^{\mu}\right)  ^{2}\left(
\bar{k}_{y}^{\mu}\right)  ^{2}=\overline{i}_{x}^{\mu}\bar{j}_{x}^{\mu}\bar
{k}_{x}^{\mu}=-1$, which conflicts with the first relation in Eq.
(\ref{confl2}), regardless of the explicit forms of the probabilities in
(\ref{local4}) and the specific values of $\overline{i}_{y}^{\mu}$, $\bar
{j}_{y}^{\mu}$, and $\bar{k}_{y}^{\mu}$. Thus, quantum-mechanical predictions
(\ref{ijk1}-\ref{ijk4}) are incompatible with locality [Eq. (\ref{local4})]
for the GHZ states (\ref{ghz}) and the incompatibility is manifested
\textit{without inequalities}. The present proof implies that the GHZ
experiment performed by Pan \textit{et al}. \cite{Pan-GHZ} actually ruled out
locality, but not realism.

Some interesting new features emerge in the present proof. First, although the
procedure for proving the conflict is somewhat similar to that used in the GHZ
argument \cite{GHZ-89,GHZ-90,Mermin,Pan-GHZ}, here we do not in any way use
the realism assumption: $\overline{i}_{x,y}^{\mu}$, $\bar{j}_{x,y}^{\mu}$ and
$\bar{k}_{x,y}^{\mu}$\ in Eqs. (\ref{def-ijk}-\ref{confl2}) should not be
confused with the corresponding local realistic predictions occurring in the
GHZ argument; they are predictions under the locality assumption. The perfect
correlations among the three qubits and the logic link leading to the
conflict, as shown in (\ref{confl2}), are recovered at the level of average
values. This is in sharp contrast to the GHZ argument and Eq. (\ref{allno}).
Thus, the conflict proved here is not of an all-versus-nothing type due to the
probabilistic nature of the locality condition (\ref{local4}).

By assuming QM, the conflict will be even stronger than that shown in
(\ref{confl2}). To see this, recall that in QM, the quantities defined in
(\ref{def-ijk}) are quantum mechanical\ predictions of the three local parties
and as such, they are constrained by the following Heisenberg-Robertson
relations $\overline{\gamma}_{x}^{\mu2}+\overline{\gamma}_{y}^{\mu2}\leq1$ and
thus, $\left\vert \overline{\gamma}_{x}^{\mu}\right\vert $, $\left\vert
\overline{\gamma}_{y}^{\mu}\right\vert \leq1$, which imply that either
$\left\vert \overline{\gamma}_{x}^{\mu}\right\vert $, $\left\vert
\overline{\gamma}_{y}^{\mu}\right\vert <1$\ or one of $\left\vert
\overline{\gamma}_{x}^{\mu}\right\vert $ and $\left\vert \overline{\gamma}%
_{y}^{\mu}\right\vert $\ is $1$ (and another will necessarily be zero). For
the former case, each of the four relations in (\ref{confl2}) cannot be valid,
while for the latter case, at most one relation in (\ref{confl2}) is
valid.\ This stronger conflict between locality and the GHZ correlations\ is
in sharp contrast to that shown in (\ref{confl2}) (as well as in the GHZ
\cite{GHZ-89,GHZ-90,Mermin,Pan-GHZ} or Stapp's reasoning \cite{Stapp-GHZ}),
where a combination of three relations conflicts the remaining one. Since here
we have used the quantum probabilities for envents in each location, this
nonlocality is of a quantum nature, i.e., quantum nonlocality.

There are also biparticle perfect correlations as used in the usual EPR
argument. Following the above discussion, one cannot obtain a similar conflict
for the perfect EPR correlations if only locality is assumed. When quantum
results (i.e., the Heisenberg-Robertson relations) are used as above, a
conflict does occurs. We anticipate this to be the essence of the EPR reasoning.

In the above \textit{gedanken} GHZ experiment, one assumes perfect
correlations and ideal measurement devices.\ In a real experiment, these ideal
requirements are practically impossible. To face this difficulty, a
\textquotedblleft local inequality\textquotedblright\ for the GHZ state is
desirable to take into account the imperfections of real experiments. For this
purpose, one can introduce the \textquotedblleft GHZ-Mermin
operators\textquotedblright\ \cite{Mermin65}
\begin{align}
M  &  =\left(  \sigma_{1x}\sigma_{2y}+\sigma_{1y}\sigma_{2x}\right)
\sigma_{3y}+\left(  \sigma_{1y}\sigma_{2y}-\sigma_{1x}\sigma_{2x}\right)
\sigma_{3x},\nonumber\\
M^{\prime}  &  =\left(  \sigma_{1x}\sigma_{2y}+\sigma_{1y}\sigma_{2x}\right)
\sigma_{3x}-\left(  \sigma_{1y}\sigma_{2y}-\sigma_{1x}\sigma_{2x}\right)
\sigma_{3y}.\nonumber\\
&  \label{m}%
\end{align}
Using the locality condition (\ref{local4}), one can prove the locality
inequality
\begin{equation}
\max\left\{  \left\vert \left\langle M\right\rangle _{LT}\right\vert
,\left\vert \left\langle M^{\prime}\right\rangle _{LT}\right\vert \right\}
\leq2, \label{lt}%
\end{equation}
which is imposed on \textit{any} local theory. As a comparison, recall that
local realistic theories satisfy \cite{Mermin65,Bell-n} the same inequality as
(\ref{lt}), i.e., $\max\left\{  \left\vert \left\langle M\right\rangle
_{LRT}\right\vert ,\left\vert \left\langle M^{\prime}\right\rangle
_{LRT}\right\vert \right\}  \leq2$. Meanwhile, realistic theories
(\textit{without the locality assumption}) predict
\begin{equation}
\max\left\{  \left\vert \left\langle M\right\rangle _{RT}\right\vert
,\left\vert \left\langle M^{\prime}\right\rangle _{RT}\right\vert \right\}
\leq4. \label{rt}%
\end{equation}
Similarly to Ref. \cite{Chen}, the inequality (\ref{rt}) can be proved by
noticing that the absolute values of each of the eight observable quantities
in Eq. (\ref{m}) are equal to or less than $1$ in a realistic theory.

However, if the local probabilities in the locality condition (\ref{local4})
are given by QM, then a quantum locality inequality, stronger than the
locality inequality (\ref{lt}), can be proved for three-qubit states $\rho$:
\begin{equation}
\left\langle M\right\rangle _{\rho,L}^{2}+\left\langle M^{\prime}\right\rangle
_{\rho,L}^{2}\leq1, \label{local-ineq}%
\end{equation}
where $\left\langle M\right\rangle _{\rho}=\mathrm{Tr}(\rho M)$. For the three
qubits in the GHZ state (\ref{ghz}) one obtains, from Eqs.\ (\ref{ghz4}),
$M\left\vert \mathrm{\Delta}\right\rangle =4\left\vert \mathrm{\Delta
}\right\rangle $, resulting in $\left\langle M\right\rangle _{\rho}%
^{2}+\left\langle M^{\prime}\right\rangle _{\rho}^{2}=16$, which conflicts the
inequality (\ref{local-ineq}). Violation of the quantum locality inequality
(\ref{local-ineq}) implies genuine quantum nonlocality. Without assuming
locality, QM\ predicts an inequality \cite{Uffink,Yu}
\begin{equation}
\left\langle M\right\rangle _{\rho,QM}^{2}+\left\langle M^{\prime
}\right\rangle _{\rho,QM}^{2}\leq16 \label{qm}%
\end{equation}
for all three-qubit states.

As we proved in Ref. \cite{whole}, the locality condition given by QM is a
necessary and sufficient condition for $N$-qubit states to be totally
separable so far as the $N$ qubits are mutually spacelike separated. Thus, the
quantum locality inequality (\ref{local-ineq}) is fulfilled by totally
separable quantum states, which are quantum mechanically local by definition
\cite{Chen}. Any three-qubit state has quantum nonlocality if it violates the
quantum locality inequality (\ref{local-ineq}). It is worthwhile to mention
that three-qubit states satisfying $8<\left\langle M\right\rangle _{\rho
,(3)}^{2}+\left\langle M^{\prime}\right\rangle _{\rho,(3)}^{2}\leq16$\ are
totally entangled (i.e., $3$-entangled), while partially entangled (i.e.,
$2$-entangled) three-qubit states fulfil $1<\left\langle M\right\rangle
_{\rho,(2,1)}^{2}+\left\langle M^{\prime}\right\rangle _{\rho,(2,1)}^{2}\leq
8$. This result is useful in detecting three-qubit quantum nonlocality (or
entanglement).%
\begin{figure}
[ptb]
\begin{center}
\includegraphics[
height=2.222in,
width=2.2096in
]%
{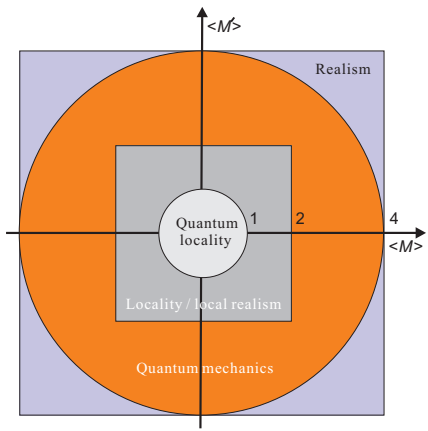}%
\caption{Four inequalities (\ref{lt}-\ref{qm}) in the $\left\langle
M\right\rangle $-$\left\langle M^{\prime}\right\rangle $ plane: The quantum
locality inequality (\ref{local-ineq}) (the inner circle with radius $1$); the
inequality (\ref{lt}) imposed by locality or local realism (the inner square);
the inequality (\ref{rt}) (the outer square); the inequality (\ref{qm}) (the
outer circle with radius $4$).}%
\label{fig}%
\end{center}
\end{figure}

The four inequalities (\ref{lt}-\ref{qm}) are plotted in the $\left\langle
M\right\rangle $-$\left\langle M^{\prime}\right\rangle $ plane (Fig. $1$). The
diagram shows that for the GHZ experiment, all quantum predictions are a
subset of the predictions of classical realism; locality alone can lead to
contradictions with the GHZ correlations. If the probabilities in the locality
condition (\ref{local4}) are quantum mechanical predictions, then stronger
contradictions can be revealed than locality alone or local realism. This
observation is similar to the results obtained in Ref. \cite{Chen} for the
Bell experiments. The experimental verification of violating the quantum
locality inequality (\ref{local-ineq})] implies a genuine quantum nonlocality
that is not masked by the realism assumption.

To summarize, we have proved that for a three-qubit system in the GHZ state
locality is in conflict with the perfect GHZ correlation. By contrast to the
GHZ theorem refuting local realism, the present proof refutes only locality
and does not in any way use the realism assumption, which cannot be ruled out
by the standard GHZ experiments. Thus, the present result reveals a genuine
(multiparticle) quantum nonlocality, i.e., \textit{quantum nonlocality beyond
Bell's nonlocality}. We have also provided inequalities that are imposed by
locality and can be maximally violated by the GHZ-entangled qubits. This is
essential for a practical GHZ experiment. Similar considerations on Hardy's
theorem \cite{Hardy} and Cabello's theorem \cite{Cabello} will be given elsewhere.

We thank Jian-Wei Pan for stimulating discussion. This work was supported by
the National NSF of China under Grant No. 10104014, the CAS and the National
Fundamental Research Program under Grant No. 2001CB309300. N.-L.L. was also
supported by the Scientific Research Foundation for the Returned Overseas
Chinese Scholars by the State Education Ministry of China and the USTC
Returned Overseas Scholars Foundation.

\end{document}